\documentclass{easychair}
\pdfoutput=1
\usepackage{subfigure}
\usepackage{prettyref}
\newrefformat{sec}{Section~\ref{#1}}
\newrefformat{tab}{Table~\ref{#1}}
\newrefformat{fig}{Figure~\ref{#1}}
\newrefformat{cha}{Chapter~\ref{#1}}
\newrefformat{app}{Appendix~\ref{#1}}

\usepackage{booktabs}

\usepackage{xspace}
\makeatletter

\newcommand{\eg}{e.g.\@\xspace}

\makeatother

\usepackage[breaklinks]{hyperref}

\begin{document}

\date{}

\title{Chip and Skim: cloning EMV cards\\ with the pre-play attack}

\author{Mike Bond, Omar Choudary, Steven J. Murdoch,\\Sergei Skorobogatov, \and Ross Anderson\\
{\tt forename.lastname@cl.cam.ac.uk}}

\institute{Computer Laboratory, University of Cambridge, UK}

\titlerunning{Chip and Skim}
\authorrunning{Bond, Choudary, Murdoch, Skorobogatov and Anderson}

\maketitle

\begin{abstract}

EMV, also known as ``Chip and PIN'', is the leading system for card payments worldwide. It
is used throughout Europe and much of Asia, and is starting to be introduced in North
America too. Payment cards contain a chip so they can execute an authentication protocol.
This protocol requires point-of-sale (POS) terminals or ATMs to generate a nonce, called
the unpredictable number, for each transaction to ensure it is fresh. We have discovered
that some EMV implementers have merely used counters, timestamps or home-grown algorithms
to supply this number. This exposes them to a ``pre-play'' attack which is
indistinguishable from card cloning from the standpoint of the logs available to the
card-issuing bank, and can be carried out even if it is impossible to clone a card
physically (in the sense of extracting the key material and loading it into another card).
Card cloning is the very type of fraud that EMV was supposed to prevent. We describe how
we detected the vulnerability, a survey methodology we developed to chart the scope of the
weakness, evidence from ATM and terminal experiments in the field, and our implementation
of proof-of-concept attacks. We found flaws in widely-used ATMs from the largest
manufacturers. We can now explain at least some of the increasing number of frauds in
which victims are refused refunds by banks which claim that EMV cards cannot be cloned and
that a customer involved in a dispute must therefore be mistaken or complicit. Pre-play
attacks may also be carried out by malware in an ATM or POS terminal, or by a
man-in-the-middle between the terminal and the acquirer. We explore the design and
implementation mistakes that enabled the flaw to evade detection until now: shortcomings
of the EMV specification, of the EMV kernel certification process, of implementation
testing, formal analysis, or monitoring customer complaints. Finally we discuss
countermeasures.

\end{abstract}

\section{The Smoking Gun}

EMV is now the leading scheme worldwide for debit and credit card payments, as
well as for cash withdrawals at ATMs, with more than 1.34 billion cards in
use worldwide.  US banks were late adopters, but are now in starting to issue
EMV cards to their customers. EMV cards contain a smart card chip, and are more
difficult to clone than the magnetic-strip cards that preceded them.

EMV was rolled out in Europe over the last ten years, with the UK being one of
the early adopters (from 2003--5). After it was deployed, the banks started to
be more aggressive towards customers who complained of fraud, and a cycle
established itself. Victims would be denied compensation; they would Google for
technical information on card fraud, and find one or other of the academic
groups with research papers on the subject; the researchers would look into 
their case history; and quite often a new vulnerability would be discovered.

The case which kicked off the research we report here was that of a Mr Gambin,
a Maltese customer of HSBC who was refused a refund for a series of
transactions that were billed to his card and which HSBC claimed must have been
made with his card and PIN at an ATM in Palma, Majorca on the 29th June 2011.
In such cases we advise the fraud victim to demand the transaction logs from the
bank. In many cases the banks refuse, or even delete logs during the dispute
process, leaving customers to argue about generalities. Some courts have recently
criticised banks for this and in the Gambin case the bank produced detailed log
data. We observed that one of the fields on the log file, the ``unpredictable number''
or UN, appeared to be increasing steadily:

{}

\begin{center}
\begin{tabular}{ccc}
\toprule
Date & Time & UN \\
\midrule
2011-06-29 & 10:37:24 & \texttt{F1246E04} \\
2011-06-29 & 10:37:59 & \texttt{F1241354} \\
2011-06-29 & 10:38:34 & \texttt{F1244328} \\
2011-06-29 & 10:39:08 & \texttt{F1247348} \\
\bottomrule
\end{tabular}
\end{center}

The UN appears to consist of a 17 bit fixed value and the low 15 bits are simply a
counter that is incremented every few milliseconds, cycling every three
minutes.

We wondered whether, if the ``unpredictable number'' generated by an ATM is in
fact predictable, this might create the opportunity for an attack in which a
criminal with temporary access to a card (say, in a Mafia-owned shop) can
compute the authorisation codes needed to draw cash from that ATM at some time
in the future for which the value of the UN can be predicted.
We term this scenario the ``pre-play'' attack.

We discovered that several ATMs generate poor random numbers, and that attacks
are indeed possible. Following our responsible disclosure policy, we informed
bank industry organisations in early 2012 so that ATM software can be patched.
We are now publishing the results of our research so that customers whose
claims for refund have been wrongly denied have the evidence to pursue them,
and so that the crypto, security and bank regulation communities can learn the
lessons. These are considerable. For engineers, it is fascinating to unravel
why such a major failure could have been introduced, how it could have
persisted undiscovered for so long, and what this has to tell us about
assurance. At the scientific level, it has lessons to teach about the nature of
revocation in cryptographic protocols, the limits of formal verification, and
the interplay between protocol design and security economics.

The rest of this paper is organised as follows. In \prettyref{sec:background},
we give the high-level background, telling the history of EMV and discussing
its effect on fraud figures overall. In \prettyref{sec:emv-overview} we give
the technical background, describing how an EMV transaction works.
\prettyref{sec:experiment} describes our experimental methods and results: how
we developed a data capture card to harvest UN sequences from ATMs, and what we
learned from examining second-hand ATMs bought on eBay.
\prettyref{sec:discussion} presents our scientific analysis: what the crypto
and security communities should take away from this, how EMV can be made more
robust, and how such failures can be made less likely in future large-scale
systems that employ cryptography for authentication and authorisation. Finally
in \prettyref{sec:conclusions} we draw some conclusions.

\section{Background}
\label{sec:background}

EMV (named after its original developers Europay, MasterCard and Visa) was
developed in the mid 1990s to tackle the developing threat of magnetic strip
card counterfeiting, where organised crime gangs with access to card
manufacturing equipment produced cloned cards using data from discarded
receipts, or skimmed surreptitiously from legitimate cards, first at
point-of-sale (POS) and later at automated teller machines (ATMs).  The payment
terminal executes the EMV protocol with the chip, which exchanges selected
transaction data sealed with a cryptographic message authentication code (MAC)
calculated using a symmetric key stored in the card and shared with the bank
which issued the card (the ``issuer''). The idea is that the bank should be
able to detect a counterfeit card that does not contain this key, and the
physical tamper-resistance of the chip should prevent an attacker from
extracting the key.

Many countries, including the UK, moved to authenticating cardholders with a
PIN rather than a signature at both POS and ATM, where previously PINs had only
been used at ATMs. The goal was to make it harder to use a stolen card. This
simultaneous introduction gave rise to the term ``Chip and PIN'' being commonly
used in the English-speaking world to refer to EMV. In layman's terms, the chip
protects against card counterfeiting, and the PIN against stolen card abuse.

\begin{figure}
\begin{center}
\includegraphics[width=14cm]{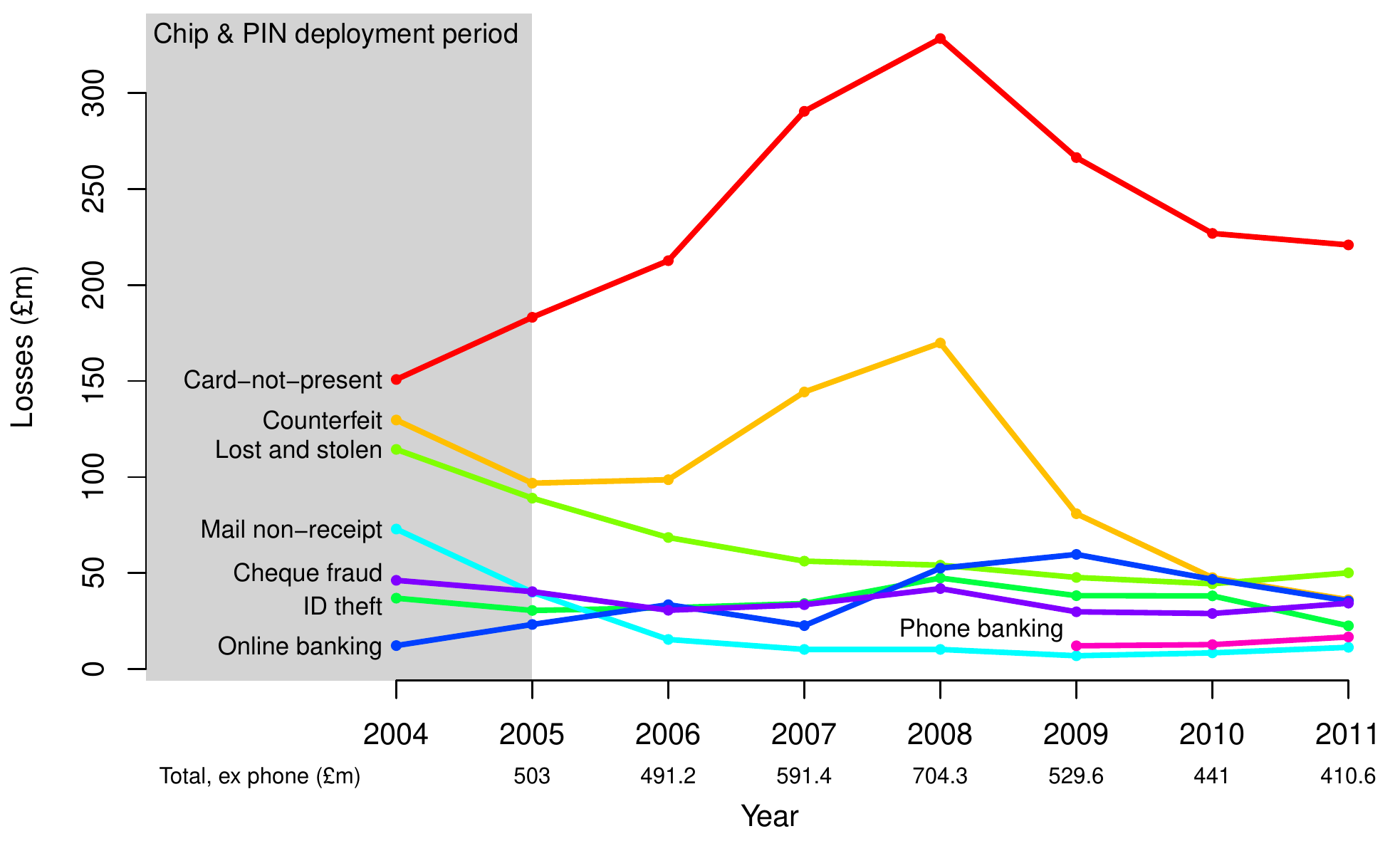}
\end{center}
\caption{\label{fig:apacs-fraud}Fraud levels on UK-issued payments cards}
\end{figure}

EMV did not cut fraud as its proponents predicted. While using counterfeit and
stolen cards did become more difficult, criminals adapted in two ways, as can
be seen from~\prettyref{fig:apacs-fraud}. First, they moved to
``card-not-present'' transactions -- Internet, mail-order, and phone-based
payments -- which remained beyond the scope of EMV.

Second, they started making magnetic-strip clones of EMV cards. There had
always been some ATM ``skimming'' where crooks put devices on ATM throats to
capture card data and record PINs; and now that PINs were demanded everywhere
and not just at ATMs, the opportunities for skimming increased hugely. The
simultaneous deployment of EMV with magnetic strip meant that fallback and
backwards-compatibility features in EMV could be exploited; for several years,
all ATMs would still accept mag-strip cards, and even once this started to be
phased out in the UK for locally-issued cards, it was still possible to use
mag-strip clones of UK cards in ATMs in the USA. This is why, soon after the
completion of the UK EMV roll-out in 2005, counterfeit fraud went up. Instead
of entering PINs only at ATMs, customers were now entering their PIN in POS 
terminals, which are much easier to tamper with~\cite{tamperterm}.

Total fraud levels were brought down following 2008 through improvements to
back-end fraud detection mechanisms which reject suspicious transactions; by
more aggressive tactics towards customers who dispute transactions; and by
reducing the number of UK ATMs that accept ``fallback'' magnetic-strip
transactions on EMV-issued cards. Fallback fraud is now hard enough to push the
criminal community to more sophisticated smart-card-based attacks.

Prior research showed that it was possible to use a stolen EMV card in a POS
device without knowing the PIN. Given a suitable man-in-the-middle device, a
crook can trick the terminal into believing that the right PIN was entered,
while the card thought it was authorising a chip-and-signature
transaction~\cite{nopin}; criminals have now gone on trial in France for
exploiting this ``no pin'' vulnerability~\cite{nopin-exploit}.

However, the ``no pin'' vulnerability does not explain the large number of
people who have contacted the authors having been refused a refund for a
fraudulent ATM transaction which they adamantly deny having made. One such case
was that of Alain Job who sued his bank for a refund, but lost after the judge 
concluded that the customer's card was probably used, not a 
clone~\cite{job-case}. In that case, the bank destroyed the log files despite 
the fact that a dispute was underway, contrary to Visa guidelines, and the 
judge warned that a court might not be so tolerant of such behaviour in the
future.

The number of such cases is unknown. The UK fraud figures quoted above only
count losses by banks and by merchants, not those for which customers are
blamed; and since the introduction of EMV, the banks have operated a ``liability
shift'' as they describe it, which means that when a transaction is disputed,
then if a PIN was used the customer is held liable, while if no PIN was used
the transaction is charged back to the merchant. This may be ideal from the
banks' viewpoint but is less so for their customers.  The 2008/2009 British
Crime Survey~\cite{acqcrime} found that 44\% of fraud victims didn't get all
their money back, despite both bank guidelines and the European Payment
Services Directive requiring that customers who have not acted negligently or
dishonestly be refunded.  Of the 44\% who were not fully refunded for their
losses, 55\% lost between \pounds 25 and \pounds 499 (\$40 to \$790) and 32\%
lost \pounds 500 or more. So there's a large gap between the banks' statistics
and those from the crime survey. We believe that the vulnerability we expose in
this paper could explain some of it.

\section{Overview of an ATM transaction}
\label{sec:emv-overview}

\begin{figure*}
\includegraphics[width=\textwidth]{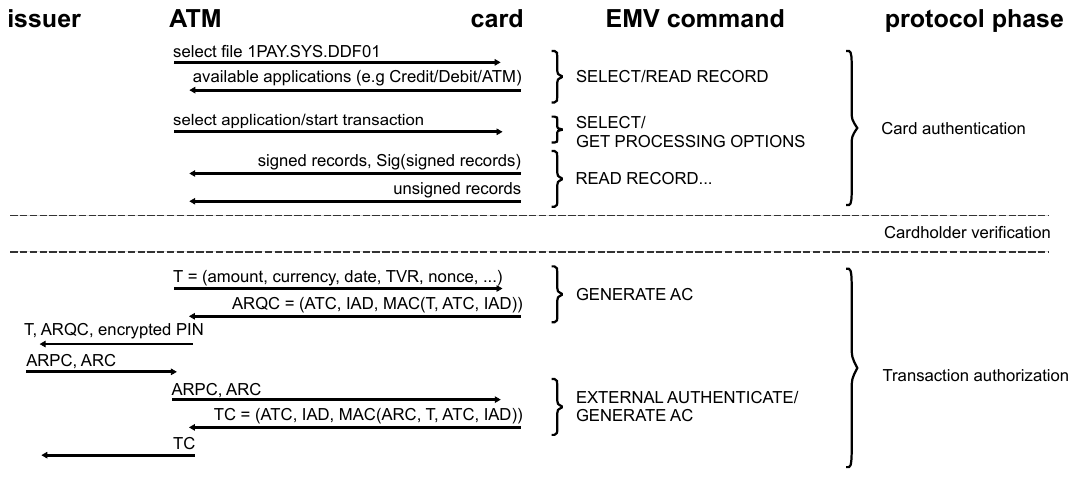}
\caption{\label{fig:protocol}Outline of an EMV transaction at ATM. Note that while the messages between card and ATM have been verified, messages between issuer and ATM may vary depending on card scheme rules}
\end{figure*}

An EMV transaction consists of three phases:

\begin{enumerate}
\item {\bf card authentication} in which card details are read and
  authenticated by the ATM or POS terminal; 
\item {\bf cardholder verification} in which the person who presents the
  card is verified whether by PIN or signature; and 
\item {\bf transaction authorization} in which the issuing bank decides whether
  the transaction should proceed.
\end{enumerate} 
The principals are the card, the ATM and the
issuer\footnote{The bank that operates the ATM (the acquirer) and the network
  that links the issuer to the acquirer are also involved in settlement,
  dispute resolution and assurance, but they do not participate in the
  authentication protocol run other than to route messages, so have been omitted
  from the discussion in this section.}.  The process is illustrated in
\prettyref{fig:protocol}.  The description below has been somewhat simplified,
and represents typical UK transaction flow. Other countries may differ
slightly, but will be substantially similar.

During card authentication, the card provides data records to the ATM, which
include the card number, start and expiry dates and which protocol options the
card supports.  The card also provides a static RSA digital signature over
selected records, which aims to prevent crooks from fabricating cards from 
known or guessed account numbers. Some cards also provide dynamic signature
generation capabilities, known as ``Dynamic Data Authentication'' (DDA).

Following card authentication, cardholder verification proceeds by signature or
PIN. In an ATM transaction the card is not involved in this verification. The customer enters their
PIN on the PIN pad, where it is encrypted and returned to the card issuer for
verification through the ATM network.

Finally, transaction authorization is carried out. The ATM sends to the card
various transaction fields: the amount, the currency, the date, the terminal
verification results (TVR -- the results of various checks performed by the
ATM), and a nonce (in EMV terminology, the ``unpredictable number'' or UN).  The
card responds with an authorization request cryptogram (ARQC), which is a
cryptographic MAC calculated over the supplied data, together with some
card-provided data including the application transaction counter (ATC -- a 16
bit number stored by the card and incremented on each transaction) and the
issuer application data (IAD -- a proprietary data field to carry information
from the card to its issuer).

The ARQC is sent by the ATM to the issuer along with the encrypted PIN. The
issuer verifies the PIN and checks the ARQC by recalculating the MAC over the
received data fields. Additional checks include whether sufficient funds are
available, that the card has not been reported stolen, and risk-analysis
software does not flag the transaction as suspicious.  Then the issuer returns
to the ATM an authorization response code (ARC) and an authorization response
cryptogram (ARPC) destined for the card.

The ARC authorises the ATM to dispense cash, which in turn passes the ARC and
ARPC also to the card. The card verifies the ARPC (which is typically a MAC over the
ARQC exclusive-or'ed with the ARC), and returns an authenticated settlement
record known as a transaction certificate (TC), which may be sent to the issuer
immediately, or some time later as part of a settlement process.

POS transactions proceed similarly, except that cardholder verification is
usually performed by sending the PIN to the card which checks it against a
stored value. Whether the PIN is verified locally or online makes no difference
to the attack discussed here. If a POS device generates unpredictable numbers
that can in fact be predicted, then it too will be vulnerable to a pre-play
attack.

\subsection{EMV pre-play protocol flaws}

The card sends an ARQC to the ATM to prove that it is alive, present, and
engaged in the transaction. The ATM relies on the issuer to verify this and
authorise the transaction.  Simply replaying an ARQC should not work, because a
competent issuer prevents replay by rejecting any transaction whose application
transaction counter it has already seen.  This prevents replay attacks but 
cannot assure the issuer that the ARQC was computed today rather than yesterday.
To ensure freshness, a nonce is used -- the unpredictable number (UN). This is 
a 32 bit field generated by the ATM.

The first flaw is that the EMV protocol designers did not think through
carefully enough what is required for it to be ``unpredictable''. The
specifications and conformance testing procedures simply require that four
consecutive transactions performed by the terminal should have unique
unpredictable numbers~\cite[test 2CM.085.00]{emvtesting2}. Thus a rational
implementer who does not have the time to think through the consequences will
probably prefer to use a counter rather than a cryptographic random number
generator (RNG); the latter would have a higher probability of failing
conformance testing (because of the birthday paradox).

The latest version of the EMV specification~\cite[Book 4, p57]{emv} offers some guidance as to how to
generate the unpredictable number, but previous versions left the algorithm
entirely up to implementers.  Even the suggested construction (hash or
exclusive-or of previous ARQCs, transaction counter and time) would not be
adequate for generating a truly unpredictable number because the ARQCs would be
zero if the ATM was rebooted and both the time and transaction counter are
predictable.  Yet if the attacker can predict an ``unpredictable number'' ahead
of time, he can harvest ARQCs from a card one day and use them at the ATM the
next.

The second flaw is that EMV does not include the identity of the terminal -- a
classic protocol mistake. While the EMV framework can support this through
designation in a list of fields to be MACed in the ARQC (the CDOL1), the standard format
developed by Visa (the version 10 cryptogram format~\cite{vis-card}) requires
only the terminal country code. The country in which the attacker will use its
skimmed data is trivial to predict in advance. The implication is that if the
attacker knows how to predict the UNs in a given make of ATM, he can harvest
ARQCs for use in any ATM of that type in a given country and at a given date in
the future.

These protocol vulnerabilities result in a ``pre-play'' attack --
authentication data are collected at one moment in time, and played to one of a
number of possible verifying parties at some later time that is already
determined when the data are harvested. The practical implementation is that a
tampered terminal in a store collects card details and ARQCs
as well as the PIN from a victim for use later that day, or the following
day, at ATMs of a given type. 

For example, in the case of the ATM in Palma that started this line of
research, the counter rolls over every three minutes, so an attacker might ask
a card in his store for twenty ARQCs at points in the 15-bit counter's cycle.
On visiting the ATM he could use his attack card to first calibrate the ATM's
counter, and then initiate transactions when the counter is expected to be at a
value for which he has a captured ARQC.

This is all very well in theory, but is it viable in practice? We
decided to find out.
 
\section{Experimental Method and Results}
\label{sec:experiment}

Pre-play attacks against EMV have been discussed theoretically before, but for
a real-world attack to work, there are many practical challenges. In this
section we describe our own approach to them: surveying for an exploitable
vulnerability, skimming data, and deploying the attack. Each stage of the
process must be completed by criminals with reasonable yield and an acceptably
low cost (including probability of being caught).

\subsection{Identifying vulnerable ATMs}

To identify vulnerable ATMs we took three approaches: analysis of log files,
collection of UNs in the field, and reverse engineering of ATMs.

\subsubsection{Analysis of log files}

We regularly investigate ATM withdrawals on behalf of customers in dispute with
their banks. In most cases the level of detail in logs provided by the bank is
low, but in a minority of cases detailed logs are handed over. The Palma
case got us started on this research track, and we have found one or two other
cases of suspicious UNs in logs.

Following our responsible disclosure of this vulnerability to the banks and
card brands, we have offered to help them analyse their log data, but have
so far received little or no feedback at all. We suggest that anyone in dispute
with a bank over ATM transactions where this vulnerability might be an explanation
should subpoena the bank's logs for analysis.

We have also discussed the vulnerability with a large online services firm, but
it turned out that they do not retain records of the UN.

We are particularly interested in collecting UN data from Italy, which is the only
country of which we are aware where UNs are routinely printed on all customer 
receipts. 

\subsubsection{Active probing of ATMs}
\label{sec:activeprobing}

Even where ATM logs are available, the timestamps have an accuracy of only a
second or so rather than a millisecond, so perhaps only grossly non-random UN
generation algorithms can be identified. For both researchers and crooks, a
better data collection approach is required. This needs to be moderately covert
as the public are aware of the problem of ATM skimming; using primitive
analysis tools repeatedly at an ATM may be a way to get arrested.

\begin{figure*}
\begin{center}
\includegraphics[width=7cm]{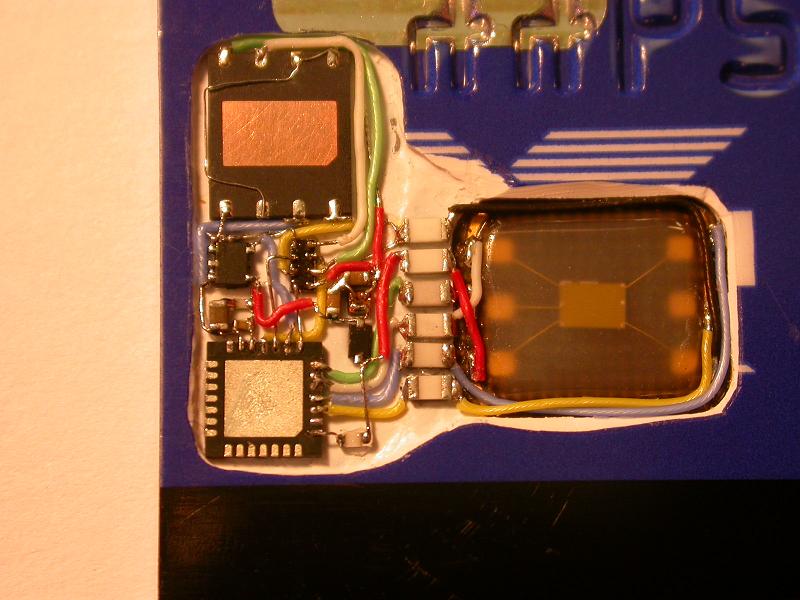}\quad\includegraphics[width=7cm]{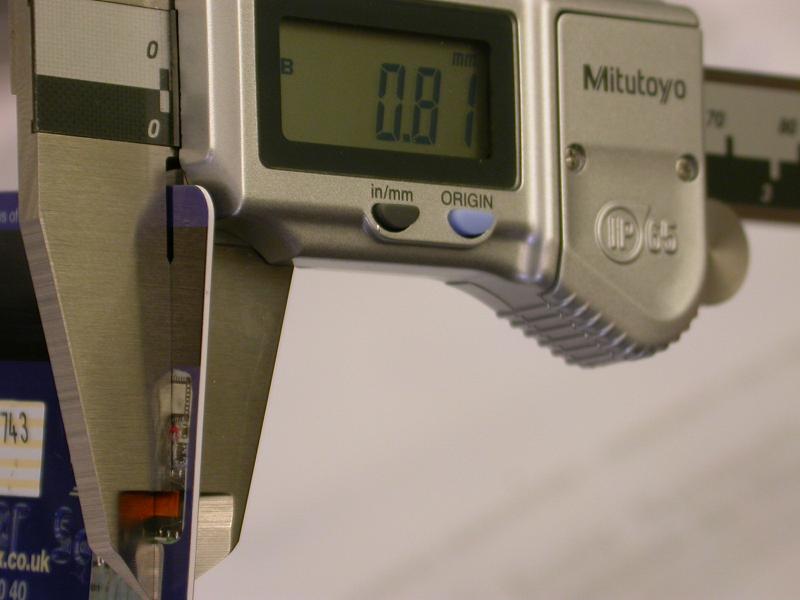}
\end{center}
\caption{\label{fig:passivemonitor} Passive monitoring card containing
  real EMV chip, with monitoring microcontroller and flash storage}
\end{figure*}

Therefore we constructed passive monitoring cards by adding a microcontroller
to existing EMV cards. (For ethical and prudential reasons we informed the
Metropolitan Police that such experiments were underway; we also consulted our
local ethics process.) Care was taken to ensure that the physical size of each
card was not modified. The card remains a valid payment card -- the transaction
flow proceeds as normal -- so it should always be accepted.  However,
it can be inserted into a variety of ATMs and POS devices without arousing
suspicion. More primitive approaches with trailing wires from the slot may
cause problems in ATMs that hold the card internally during reading.
\prettyref{fig:passivemonitor} shows a payment card adapted with our
circuitry.

Other possible monitoring equipment includes wireless relay cards transferring
data to a card outside, a wired card adapted to be compatible with ATM card
slots, an overlaid shim glued atop a thinned-down existing card, or an
ultra-simple shim consisting simply of an antenna suitably connected to the
card data line (which we could observe using ``TEMPEST'' techniques).

In the case of POS terminals, sales assistants are often briefed to turn away
during PIN entry and avoid handling the customer card. Thus existing monitoring
tools such as the Smart Card Detective~\cite{scd} have been proven suitable for
surreptitious use with a hidden wire running up the experimenter's sleeve.
We have used this tool to analyze unpredictable numbers from a POS terminal
close to our offices, having the agreement of the POS owner.

For each ATM investigated we harvested between five and fifty unpredictable numbers by
performing repeated balance enquiries\footnote{It seems all transactions at ATM
  are authenticated by EMV protocol runs, but some with a zero withdrawal
  amount.} and then a small cash withdrawal. The use of balance enquiries
minimises the number of withdrawals on the card, as sudden repeated withdrawals
might trigger a fraud detection system and cause the card to be retained. Such
cards cost a few hundred pounds in component and labour costs so it is
desirable to avoid their being captured by ATMs.

\subsubsection{Reverse engineering ATM code}
\label{sec:reverse-engineering}

\begin{figure}
\begin{center}
\includegraphics[height=6cm]{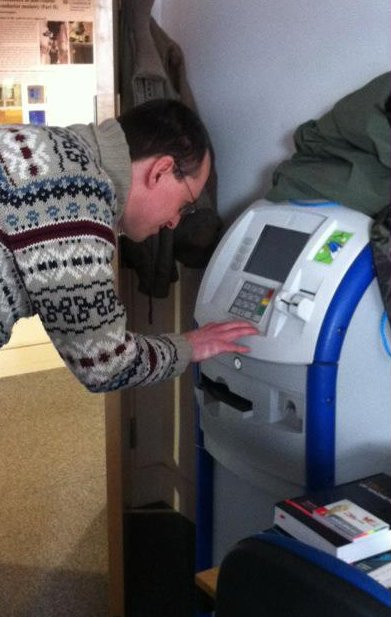}\quad%
\includegraphics[height=6cm]{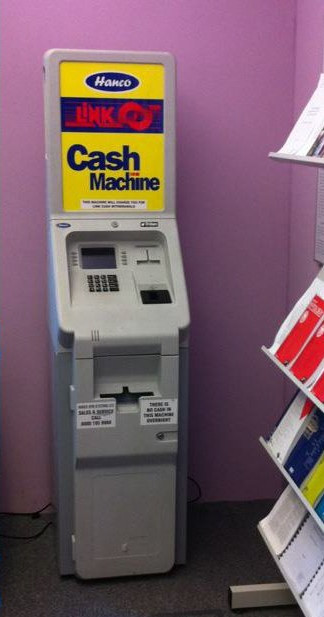}
\end{center}
\caption{\label{fig:atms} ATMs acquired for reverse-engineering}
\end{figure}

We were aware that black-box analysis of terminals through looking at lists of
UNs would not tell the full story, so we acquired some ATMs for analysis.
\prettyref{fig:atms} shows EMV-enabled NCR and Hanco/Triton ATMs acquired via
eBay for \pounds 100 each. Some of these had been in recent service, and some
were out of service, having only been used for development. Barnaby
Jack~\cite{jack} describes how second-hand ATMs can be brought back into
service easily by simply phoning for a repairman.

Reverse engineering a functioning and captive ATM can combine the best of
black-box analysis with detailed work on the algorithms and also has the
potential to expose weak pseudo-random number generators such as the C \texttt{rand()}
function whose output might look acceptable to black-box analysis but is
entirely predictable from a couple of recorded samples.

\begin{figure}%
\centering
\parbox[b]{7cm}{
 \centering
 \subfigure[][Extracting disk image from NCR ATM]{%
  \label{fig:ncr-extract}
  \includegraphics[height=11.05cm]{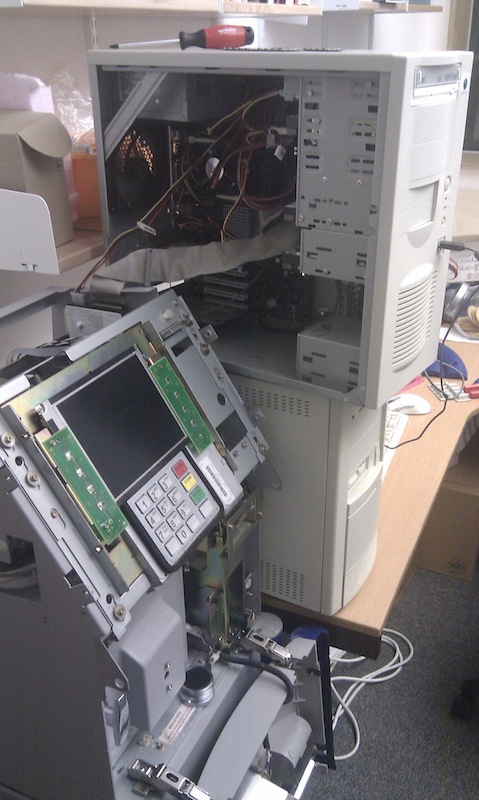}
 }
}\quad%
\parbox[b]{6.5cm}{
 \centering
 \subfigure[][Board with DES chip from Triton ATM]{%
  \label{fig:triton-desboard}
  \includegraphics[width=6.2cm]{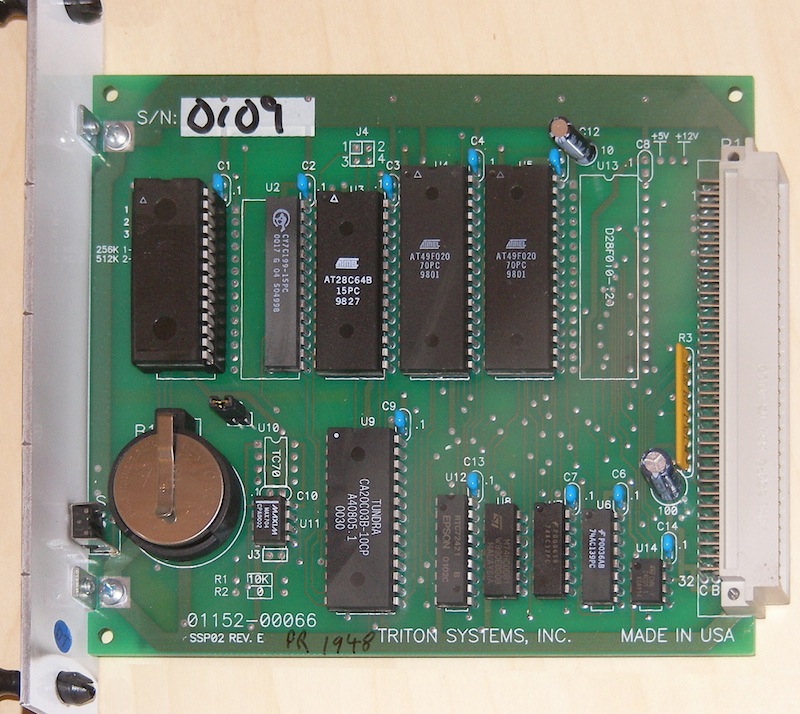}
 }
 \subfigure[][CPU board from Triton ATM]{%
  \label{fig:triton-mainboard}
  \includegraphics[width=6.2cm]{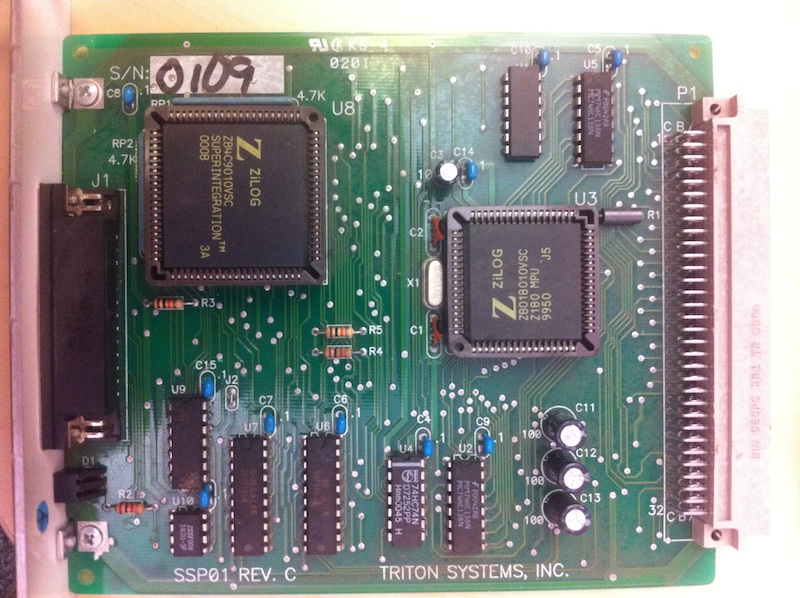}
 }
}
\caption{Detail of hardware reverse engineering}
\label{fig:atm-boards}%
\end{figure}

We have yet to confirm the UN generation algorithm in any of the ATMs.
Analysis has been complicated by the obsolete architectures and our work is
ongoing. One ATM was running OS/2 (see \prettyref{fig:ncr-extract}), and
another on primitive hardware based on the Zilog Z180 CPU (see
\prettyref{fig:triton-mainboard}).  We identified the manufacturer of the EMV
kernel from information inside the ATM, and documentation on their
website~\cite{creditcall-emv-docs} indicates that the EMV kernel requires
seeding with an external source of randomness. Hardware analysis revealed
presence of a dedicated crypto chip implementing DES (see \prettyref{fig:triton-desboard}) and we theorise also containing
a hardware random or pseudo-random number source.
Currently we are confident that each byte of the unpredictable number is
independently generated from an off-CPU resource. This would either be the
DES chip, a real-time clock (also present as a separate chip) or possibly
the smart card control unit which is a MagTek board accessed via a serial interface.

At the outset we believed that older, primitive platforms would be less likely
to have a strong source of randomness than modern platforms in all cases.
However our broader research across ATM and POS indicates a subtly different
conclusion. Entirely modern platforms are likely to call the typical OS
resources for random number generation, which nowadays are relatively strong.
Meanwhile legacy platforms may have either strong or very weak randomness
depending on whether this issue was thought about by the designers at the time.
Thirdly, legacy platforms which have been ported to more modern environments
are most likely to have weak randomness as during the porting the random number
generate custom call on the legacy platform is simply mapped across to the
easiest standard library call, such as the C \texttt{rand()} function.  In
summary it is as important to consider the lineage of the ATM or POS software
as it is to consider the current platform when estimating likelihood of
vulnerability.

\subsection{Analysing the RNG}

In \prettyref{sec:activeprobing} we described our own approaches to
data collection. Using this approach we collected data to analyse the
RNGs in EMV devices in our local area. We performed more than 1\,000
transactions across 22 different ATMs and five POS terminals. We were
successful at locating ATMs with weak RNGs, but attackers need to go
further and identify which specific UNs are most likely to occur at a
predictable future time. There are three broad classes of ineffective
RNG to consider:

\begin{itemize}

\item {\bf an obviously weak RNG algorithm}. This includes using counters or
  clocks directly as the UN, homegrown algorithms which combine obvious
  transaction data, and severe programming errors which cause the state-space
  of a better algorithm to be limited (\eg casting down to the wrong integer
  size, or submitting four BCD coded random bytes rather than four truly random
  bytes);

\item {\bf a simple RNG with little or no seeding}. There are many flavours,
  from a linear congruential generator, through encryption of the clock, to
  more messy schemes where we may find some fixed bits and some bits that
  cycle, or where a state machine starts off appearing random but ends up in a
  tight loop cycling through just a small number of values. From an embedded
  systems standpoint the typical options are the C standard library
  \texttt{time()} and \texttt{rand()} calls, neither of which have 
  unpredictable outputs from a cryptographic point of view;

\item {\bf an RNG that can be put into a predictable state}. The simplest
  failure mode is a strong RNG fed by a weak source of randomness that's
  restarted on power-up, so an attacker can force an outage or follow the
  replenishment crew. There are also systems drawing noise from an
  untrustworthy source, such as when an RNG uses data from previous
  transactions. The attacker could insert a card which seeds known values, or
  temporarily spoof the authorisation response from the bank, to push the RNG
  into a predictable state.
\end{itemize}

\prettyref{tab:untable} shows a selection of data collected from
various ATMs falling broadly into the first category of ineffective
algorithms. ATM1 and ATM2 contain a typical characteristic, which we
denote \emph{characteristic C}, where the high bit and the third nibble
of each UN are always set to zero. 11 of 22 ATMs we looked at
exhibited this characteristic. Our current levels of data allow us to
prove a non-uniform hypothesis on the data from most of these 11 ATMs
with a very good significance level. \prettyref{tab:charc} shows two
ten-transaction sequences from an ATM where the characteristic was
proven. However further analysis beyond confirming this characteristic
has not yielded statistically significant results yet. ATMs of wildly
different ages and containing different operating systems exhibited
characteristic C, so we believe it to be an artifact of a particular
EMV kernel post-processing an RNG source rather than of the RNG source
itself.

\begin{figure}
\begin{center}
\begin{tabular}[t]{rlrl}
\toprule
\multicolumn{2}{c}{SRC2 EXP6} &
\multicolumn{2}{c}{SRC2 EXP6B} \\
\midrule
0 & {\tt 77028437} & 0 & {\tt 5D01BBCF} \\  
1 & {\tt 0D0AF8F9} & 1 & {\tt 760273FE} \\
2 & {\tt 5C0E743C} & 2 & {\tt 730E5CE7} \\
3 & {\tt 4500CE1A} & 3 & {\tt 380CA5E2} \\
4 & {\tt 5F087130} & 4 & {\tt 580E9D1F} \\
5 & {\tt 3E0CB21D} & 5 & {\tt 6805D0F5} \\
6 & {\tt 6A05BAC3} & 6 & {\tt 530B6EF3} \\
7 & {\tt 74057B71} & 7 & {\tt 4B0FE750} \\
8 & {\tt 76031924} & 8 & {\tt 7B0F3323} \\
9 & {\tt 390E8399} & 9 & {\tt 630166E1} \\
\bottomrule
\end{tabular}
\end{center}
\caption{\label{tab:charc} Ten transaction sequences from a single ATM}
\end{figure}

We suspect a number ATMs and POS will simply be using the C standard
library \texttt{rand()} function, and are undertaking analysis using
techniques based on spectral tests. Such analysis is complicated by
the unknown levels of post-processing of the RNG: for example, we know
in the case of one EMV library that each byte of the unpredictable
number is sampled separately from the RNG -- hence a modulo 256 or a
type-cast is almost certainly post-processing the output. Multiple
calls to the RNG to produce one UN is on the one hand disadvantageous
in that fewer bits are available to detect state per sample, but
making four consecutive calls in a row for one UN reduces the
potential interference from other services within an ATM also making
calls as part of the transaction process.

The third category could possibly be spotted from empirical analysis
but are best detected with reverse-engineering. In
\prettyref{tab:untablepos} we show a list of stronger consecutive
unpredictable numbers retrieved from a local POS terminal.  Even in
this case the first bit appears to remain 0, which might suggest the
use of a signed integer.

\begin{figure}
\centering
\subfigure[][From Various ATMs]{
\label{tab:untable}
\parbox[t]{7cm}{
\centering
\begin{tabular}[t]{rlrl}
\toprule
\multicolumn{2}{c}{Counters} &
\multicolumn{2}{c}{Weak RNGs} \\
\midrule
ATM4 & {\tt eb661db4} & ATM1 & {\tt 690d4df2} \\
ATM4 & {\tt 2cb6339b} & ATM1 & {\tt 69053549} \\
ATM4 & {\tt 36a2963b} & ATM1 & {\tt 660341c7} \\
ATM4 & {\tt 3d19ca14} & ATM1 & {\tt 5e0fc8f2} \\
 &&& \\
ATM5 & {\tt F1246E04} & ATM2 & {\tt 6f0c2d04} \\
ATM5 & {\tt F1241354} & ATM2 & {\tt 580fc7d6} \\
ATM5 & {\tt F1244328} & ATM2 & {\tt 4906e840} \\
ATM5 & {\tt F1247348} & ATM2 & {\tt 46099187} \\
 &&& \\
 && ATM3 & {\tt 650155D7} \\
 && ATM3 & {\tt 7C0AF071} \\
 && ATM3 & {\tt 7B021D0E} \\
 && ATM3 & {\tt 1107CF7D} \\
\bottomrule
\end{tabular}
\vspace*{2ex}
}
}
\quad
\subfigure[][From local POS terminal]{
\label{tab:untablepos}
\parbox[t]{4cm}{
\centering
\begin{tabular}[t]{rl}
\toprule
\multicolumn{2}{c}{Stronger RNGs} \\
\midrule
POS1 & {\tt 013A8CE2} \\
POS1 & {\tt 01FB2C16} \\
POS1 & {\tt 2A26982F} \\
POS1 & {\tt 39EB1E19} \\
POS1 & {\tt 293FBA89} \\
POS1 & {\tt 49868033} \\
\bottomrule
\end{tabular}
\vspace*{2ex}
}
}
\caption{Categorised unpredictable numbers}
\end{figure}

Once UN generation is adequately understood, the attackers figure out
what UNs to collect in order to maximise the yield in the subsequent
cash-out phase. The result is a target ATM profile which is sent
together with intended withdrawal amounts, country code and date to
the gang tasked with harvesting the ARQCs. Once a vulnerable ATM using the
known RNG is identified, and the attack flow can proceed further.

\subsection{Harvesting the data}

Given temporary access to an EMV card, whose holder is prepared to enter the
PIN, and a range of possible unpredictable numbers to be harvested, the crook
programs his evil terminal to read the static data from the card and call
\texttt{GENERATE AC} to obtain an ARQC and TC for each possible UN.  This
process could be performed by a dedicated device, or by a tampered point of
sale terminal, vending machine, or ATM. The criminal could tamper with an ATM 
or point-of-sale terminal to perform these operations after (or instead of) a 
legitimate transaction.  Criminals have already shown the ability to tamper 
with equipment on an industrial scale and with great sophistication.

For each card a set of ARQCs can be harvested, perhaps many dozens. The only
limitation is the time that the card can legitimately be left in a sabotaged
POS while the customer believes that the machine is waiting for
authorisation. Thirty seconds is the standard authorisation time limit; this
might allow for more than 100 transactions to be skimmed.

\subsection{Cashing out}

To deploy the attack against an RNG which is a fast-moving counter such as we
have observed, the attacker needs to start the ATM transaction at precisely the
right moment. For a counter ticking hundreds or even thousands of times a
second, it is impractical to synchronise merely through timed insertion of the
card into the machine. A special smart card is therefore required which observes
the counter and uses an on-board clock to decide when to initiate the relevant
parts of the protocol.  Smart cards are allowed to delay processing responses
almost indefinitely using the \emph{request more time} signal (i.e. sending
byte \texttt{0x60}), and timely insertion to the nearest second will mean that
the card should never need to delay more than a few hundred milliseconds.

This requires a card with an on-board clock which will keep working even in the
absence of external power. We are developing a 16 bit microcontroller with an
on-board real-time clock (RTC), powered by a capacitor when no power is supplied.
The RTC is used to synchronise an internal high resolution timer once the card
is powered up, and waits the necessary amount of time until the ATM arrives at
the appropriate step in the EMV protocol where the unpredictable number is
sampled.

The feasibility of this attack is affected by the speed of the timer, the
process by which the ATM samples the timer, and the synchronisation resolution
of the card. However there are straightforward ways to relax the timing
requirements. The attackers simply harvest a set of transactions with
consecutive unpredictable numbers, and the attack card makes its best attempt
at synchronisation. Once the card sees the unpredictable number returned by the ATM
it looks this up in an internal lookup table. If the UN is not present, the
card can feign failure.  So if ten transactions are harvested from the skimmed
card, the timing requirements can perhaps be relaxed by a factor of ten as
well.

In the case of ATMs employing stateful predictable pseudo-random RNGs, none of
this intricacy is necessary -- the attacker simply samples the previous few
unpredictable numbers and can then predict the next one. In any case,
synchronisation technology can be developed and tested entirely offline against
captive ATMs without any need to interact with the real payment network.

\subsection{Implementation and evaluation}

\begin{figure}
\begin{center}
\includegraphics[height=10.5cm]{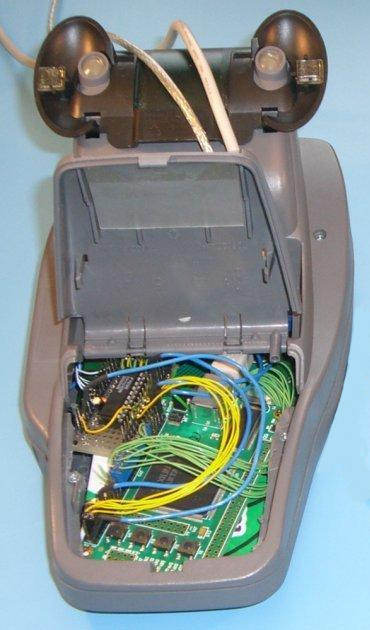}\quad%
\includegraphics[height=10.5cm]{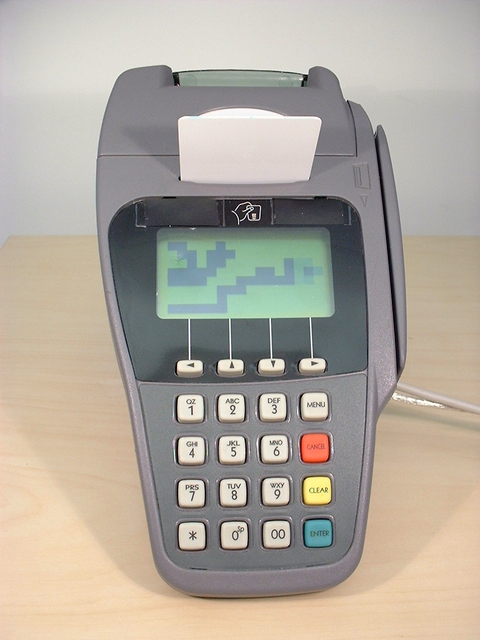}
\end{center}
\caption{\label{fig:tetris}Modified Chip and PIN terminal, playing Tetris}
\end{figure}

We have constructed proof-of-concept implementations for all stages of the
attack.  As discussed above, we modified a bank smart card for data collection
to identify ATMs with poor UN generation.  To collect card data we have
implemented a Python EMV terminal implementation and modified an EMV terminal
to collect card data, as shown in \prettyref{fig:tetris}.  To carry out the
attack we implemented a cloned card on the ZeitControl BasicCard platform.

We used test cards with known ARQC-generation keys (UDK) to prove the viability of the attack at a
protocol level. Our proof consists of an indistinguishability experiment; we
take two test cards A and B loaded with the same ARQC-generation keys,
initialised with the same ATC and handled identically. We use our skimming
trace to harvest data from card A and then program it on to a ``pre-play
card''. We then compare traces between the pre-play card version of card A and
the real card B, and observe that they are identical. This means that at a
protocol level it is impossible for the ATM to distinguish between the real and
pre-play cards. In detail the flow is as follows:

\begin{enumerate}

\item two transactions performed on card A 

\item two transactions performed on card B

\item traces of transactions compared, \texttt{GENERATE AC} responses
  confirmed the same, proving both cards have the same cryptographic
  keys and are generating the same cryptograms (they are identical)

\item two ARQCs skimmed from card A

\item pre-play card programmed with data from data collected from card A

\item two transactions performed on card B

\item two transactions performed on pre-play card

\item traces of transaction compared and shown to be identical,
  confirming that pre-play card is indistinguishable from card B

\end{enumerate}

\subsection{Limitations and Defences}

The limitations of a pre-play attack are: 
\begin{itemize}
\item The country of attack must be chosen in advance
\item The dates of attack must be chosen in advance
\item The amount must be chosen in advance
\end{itemize}

We assume that the unpredictable number is known ahead of time (either due to
full prediction of a pseudo-random RNG, or due to waiting until the appropriate
moment to sample a looping time-based counter).  It is not necessary to know
the terminal ID of the ATM, or time of transaction, as these are rarely (if at
all) requested by card and are not included in the generation of the ARQC.  The
cloned card can be used in any vulnerable ATM which shares the same country
code.

The simplest fix is a cryptographically secure random number generator. The UN
field is only 32 bits, and so an attacker who could collect approximately
$2^{16}$ ARQCs from a card could get a decent probability of success with
$2^{16}$ transactions at an ATM. This is not a realistic concern as an EMV card
should disable itself after $2^{16}$ transactions, and carrying out $2^{16}$
transactions at an ATM at 20 seconds each would not only take more than a day
but should rapidly propel the machine to the top of FICO's watch list.

The problem here is that fixing the random number generator is a matter for
acquiring banks, ATM vendors, merchants and POS terminal suppliers, while the
cost of fraud falls on the issuing banks and the customers. Hopefully this 
article will reduce the likelihood of risk being dumped unfairly on customers,
but what can an issuing bank do?

If an attacker requests many ARQCs from a card, the issuer may notice gaps in
the ATC sequence. Issuers should probably reject online transactions where the
ATC is lower than the highest ATC seen from that card, which would limit the
attack window to the next genuine card use. For offline transactions, however, this
cannot be done because there might be re-ordering of cryptograms.

One limitation of the skimming processes is that the \texttt{EXTERNAL
  AUTHENTICATE} call (which happens during the transaction authorization, see
\prettyref{fig:protocol}) cannot be made, as the ARPC cannot be generated
without the issuer's involvement.  This does not impair the card's ability to
generate the ARQC (which happens before \texttt{EXTERNAL AUTHENTICATE}), but it
might allow the attack to be detected by an issuer who examines the TC.  The
IAD field in the TC is not covered by the EMV specification, but additional
standards defined by Visa~\cite{vis-card}, commonly implemented by cards, do go into more
detail.  A pair of bits in the IAD indicates whether \texttt{EXTERNAL
  AUTHENTICATE} has been performed and whether it succeeded.  Although this is
not suitable for preventing the attack (because the TC is only sent to the
issuer once the ATM has completed the transaction), it could allow detection
later.

Another approach for increasing the difficulty of the attack is to force the
card to commit to the value of the ATC before the ATM presents the UN to the
card.  This is possible without having to modify cards, because a mandatory
feature of EMV is that the \texttt{GET DATA} command retrieves the current ATC.
If the pre-play card were able to exactly predict the value of the UN in a
transaction, being forced to choose an ATC would not affect the difficulty.
However, it would prevent the card from searching a list of available ARQCs and
finding one that matches.

One set of non-defences are the public-key authentication features of EMV.  The static
digital signature (Static Data Authentication -- SDA) included on the card can
be trivially copied to the pre-play card.  However, by examining records of
transactions we discovered that the terminal verification results (TVR) field
sent to the card during transaction authorization indicates that this digital
signature was not verified.  The decision not to check the digital signature
could have been made by ATM manufacturers to save the time needed to verify the
signature on the low-end CPUs in some ATMs (see
\prettyref{sec:reverse-engineering}), and the maintenance costs of updating the
root certificates, because counterfeit cards should be detected
during transaction authorization.

Even the public-key challenge-response protocol of EMV (Dynamic Data
Authentication -- DDA) would not adequately protect terminals from attack. If
DDA were commonly used by ATMs (or the attack is fielded at a point-of-sale
terminal) the signature response to the \texttt{INTERNAL AUTHENTICATE} command
can be recorded and replayed just as the ARQC is.  In our POS terminal tests the
unpredictable number sent by the terminal to the card in the \texttt{INTERNAL
  AUTHENTICATE} command is the same as for the \texttt{GENERATE AC} command.
However this gets us into the territory of what the acquiring bank might do, if
forced to by changes in card scheme rules or by legal precedents. We will return
to this question in the next section.

\section{Discussion}
\label{sec:discussion}

The potential vulnerability of EMV to a poor random number generator was
discussed in the abstract by Murdoch~\cite{digital-evidence}.  Markettos and
Moore~\cite{markettos} additionally explored how otherwise secure true random
number generators could be manipulated to produce more deterministic output.
But this paper is the first work to show that poor random number generators
exist in the wild, that they have been implicated in fraud, how they can be
exploited, and that the EMV specification does not test adequately for this
problem.

The exploit scenario described in this paper might be viewed as a variant of
the relay attack, which was explored in the context of EMV by Drimer and
Murdoch~\cite{enemies-close}. But there, the relay attack required real-time
bi-directional communication with the genuine card; the genuine card had to be
under the control of the attacker while the attack was taking place. This makes
it hard to deploy; the best attack we can think of is to have a false terminal
such as a parking meter to attract cardholders, communicating with a crook who
waits with the connected false card near an ATM. We do not know of this being
deployed in practice (though we've heard rumours). Another variant of the relay
attack is the no-PIN attack where a man-in-the-middle device tricks the
terminal into accepting a transaction after the wrong PIN was entered; that
also works in real time. That has been deployed, and crooks have been
prosecuted for it; but so far the losses appear to of the order of a million
Euros, and from one or two incidents. In contrast, delays of days to weeks
would be possible with the attack described in this paper, therefore making it
much more feasible to industrialise.

In other respects, the pre-play attack could be seen as a kind of card cloning.
We have already seen fake magnetic strip cards based on either the magnetic
strip of the genuine card, or the copy of the magnetic strip data stored on
the chip of some EMV cards. Another approach is the ``YES-card'' where the
static data from a chip is copied to a cloned chip card.  If the transaction
can be kept offline (\eg by keeping it below the ``floor limit''), the fact
that such a card cannot produce a valid ARQC or TC will not prevent the
transaction, but as the YES-card is responsible for verifying the PIN, it can
be programmed to accepted any PIN. The pre-play attack is more powerful in some
respects as it works for online transactions, and less in others as the
transaction parameters must be known in advance. Crucially, the pre-play attack will work in ATMs
while a YES-card won't (a typical YES-card attack involves buying cigarettes 
for resale, which is less convenient than stealing cash directly).

One might imagine that much more fraud could be committed with a fully cloned
card containing a copy of the ARQC-generation keys than with a card containing
pre-play data. However even a full clone will have its own ATC which will start
to diverge from that of the real card and in due course be detectable. So a 
full cloning attack might be not that much more powerful in practice than a
pre-play attack.

\subsection{Attack variants}

Even if the UN generation algorithms are patched, a number of powerful attack
variants may make pre-play attacks viable for years to come.

\begin{itemize}

\item \textbf{Malware}. There are already numerous cases of malware-infected
  ATMs operating in Eastern Europe and depending on the internal architecture
  of the ATM it may be easy for such malware to sabotage the choice of UN. In
  fact one bank suggested to us that the ATM that kicked off this whole
  research project may have been infected with malware.

\item \textbf{Supply chain attacks}. Such attacks have already been seen
  against POS terminals in the wild, and used to harvest magnetic strip data.
  So it is feasible that a criminal (or even a state-level adversary) might
  sabotage the RNG deliberately, either to act predictably all the time, or to
  enter a predictable mode when triggered via a covert channel. A suitably
  sabotaged RNG would probably only be detected via reverse engineering or
  observation of real world attacks.

\item \textbf{Collusive merchant}. A merchant might maliciously modify their
  EMV stack to be vulnerable, or inject replayed card data into the
  authorisation/settlement system. He could take a cut from crooks who come to
  use cloned cards at their store, or just pre-play transactions directly. In
  the UK, there was a string of card cloning attacks on petrol stations where a
  gang bribed store managers to look the other way when PIN pads were tampered
  with and monitoring devices inserted into network connections; exactly what
  you need to deploy a pre-play attack.

\item \textbf{Terminal cut-out}. A variant is the terminal cut-out or bypass is
  where the transaction stream between the merchant terminal and the acquirer
  is hacked to misreport the unpredictable number when triggered by a
  particular signal (\eg a particular account number or a known ARQC). This
  transaction data stream is not normally considered sensitive within the
  threat model and can be altered at will by merchant software. The attackers'
  card performing the replay can then use any UN for which it has an ARQC, and
  the true random UN made up by the terminal will never see the light of day.
  This is hard to block: there is no provision in currently deployed EMV cards
  for the terminal to confirm that its choice of UN was correctly included in
  the cryptographic MAC. The terminal cut-out could be implemented in malware
  (and there's evidence of bank botnets looking for POS devices), or in a
  merchant's back-end system (we have evidence of merchants already tampering
  with transaction data to represent transactions as PIN-verified when they
  were not, so as to shift liability).

\item \textbf{UN modification in the network}. A man-in-the-middle device
  between a POS device and the acquiring bank, perhaps at a network switch,
  would also be a good way to deploy such an attack.  This could be an
  attractive way to attack merchants that process high-value transactions, such
  as jewelers or investment firms, who might guard their premises and take care
  of their POS equipment yet still fall to a targeted attack. A pre-play attack
  would be much harder to detect than old-fashioned attacks that just convert
  deny authorisation messages into approve messages.

\end{itemize}

Perhaps the main takeaway message is that an attacker who can subvert a
merchant's premises, get access to his terminal equipment (even before it is
purchased), or get control of his network connection, can do transactions that
are indistinguishable from card cloning to the bank that issued the EMV card --
even if full card cloning is physically impossible. The EMV attack surface is a
bit bigger than one might think, especially once crooks learn how to manipulate
the protocol.

\subsection{EMV protocol issues}

The key shortcoming at the EMV protocol level is that the party depending upon
freshness in the protocol is not the party responsible for generating it. The
issuing bank depends on the merchant for transaction freshness. The merchant
may not be incentivised to provide it, may not be able to deliver it correctly
due to lack of end-to-end authentication with the issuer, and might even be
collusive (directly or indirectly). This is somewhat outside the terms of
reference of traditional academic protocol analysis.

Recently there has been some formal analysis of EMV, but this flaw was not
discovered~\cite{emvformal}.  One reason is that the UN was modelled as a
fresh nonce, even though this is not required by the EMV specification (this
omission is understandable given that the actual specification of the UN is
buried on p.1498 in an annex to the EMV specifications, totalling over 4\,000
pages).  The other is that the issuer and terminal are modelled as the same
individual, whereas in reality the relying party is the issuer and has only
limited control over the terminal behaviour. 

Let's consider the EMV protocol in the traditional academic framework. The
protocol might be idealised as (where A is the ATM, B is the issuer, and C is
the card):

\vspace{2ex}

\begin{tabular}{ll}
$A \longrightarrow C: $&$ N, V, T$\\
$C \longrightarrow A: $&$\{N, V, T\}_{KCB}$\\
$A \longrightarrow B: $&$\{A, \{N, V, T\}_{KCB}\}_{KBA}$\\
$B \longrightarrow A: $&$\{A, N\}_{KBA}$
\end{tabular}

\vspace{2ex}

An analysis using the BAN logic~\cite{Burrows90alogic} would note that
$\mathit{KCB}$ is a good key for communicating between the card and the bank,
so the bank knows that the card once said $N$, $V$ and $T$; if it concludes
that $N$ is also fresh, then it will infer that the card said all this in the
current epoch. However $N$ is not actually the card's nonce $NC$, but the
terminal's nonce $NT$, and we can't infer anything once we formalise it this
carefully.

It is well known that the assumptions used in the 1970s by the pioneers of
authentication were undermined by later ``progress''. The Needham-Schroeder
protocol~\cite{Needham:1978:UEA:359657.359659}, famously has a ``bug'' in that
the protocol can stall for an extended period of time between the third and
fourth messages, with the effect that old session keys once compromised cannot
be revoked. Needham and Schroeder defended themselves by pointing out that
their paper had quite openly assumed that principals executed the protocol
faithfully; therefore such behaviour was {\it a priori} excluded from their
model.  Our modern world of equipment that fails from time to time, and where
life is spiced by the occasional malicious insider, requires us to be more
careful with revocation.

In exactly the same way, the deployment of a system like EMV across an
ecosystem with hundreds of vendors, thousands of banks, millions of merchants
and billions of cards requires us to be much more careful about who the
principals are, and the incentives they have to execute their tasks
competently. Indeed, one of the new realities of the EMV world is that
merchants and banks may be hostile parties in the payment system, thanks to
tussles over payment transaction charges and chargebacks. There have been large
lawsuits between major retailers and payment networks, and we are aware of
cases where merchants deliberately falsify record data (e.g. by claiming that
transactions were PIN-verified when they were not) so as to push fraud costs to
the bank and reduce chargebacks.

So if issuing banks cannot trust merchants to buy terminals from vendors who
will implement decent random number generators, what can be done? The protocol
specialist will say that randomness must be generated by the party that relies
on it; so the terminal should request a nonce from the issuing bank before
commencing the transaction. This would take a long time to implement and impose
significant time and financial penalties as it requires an extra message
round trip for each authorisation.

A cheaper alternative might be a rule that, in the event of a transaction
dispute, it would be the responsibility of the acquiring bank to demonstrate
that the unpredictable number was properly generated. The terminal equipment
might support audit in various ways, such as by using a generator which
encrypted an underlying sequence that is revealed after the fact, and locked to
the transaction log to establish time limits on possible pre-play tampering.
However, this would not be entirely trivial; secure storage of audit data in
the terminal is a new problem and creates new opportunities for attack.

\subsection{Evidential issues in dispute}

Viability of the pre-play attack has significant legal ramifications. It can no
longer be taken for granted that data in a logged transaction was harvested at
the time and place claimed, which undermines the reliability of evidence in
both civil and criminal cases. To show that a given transaction was made by a
particular card, it is now necessary to show that the random number generator
on the ATM or POS was sound.

From the point of view of an issuing bank in dispute with a customer, this
attack greatly complicates matters. The bank cannot just rely on its own log
data -- it must collect data from a third party (the ATM operator) to prove
that the ATM was not infected with malware; that the random number generator
was not vulnerable due to either design failure or a supply chain attack; and
that the logs at the acquirer match those kept at the terminal itself. A mere
one-off certification for a class of EMV kernel does not come close to
discharging this burden. There may be practical matters in incentivising the
acquiring bank to cooperate with the issuer, especially in international cases.

Under existing Visa guidelines, logs should be retained in case of dispute. Yet
in recent cases we have dealt with, logs were routinely destroyed after 90 or
180 days regardless of whether a dispute was in progress. So the industry
already cannot cope with dispute resolution based on issuer logs; and given
that some of the disputes we're already seeing would require scrutiny of
acquirer and ATM operator systems, dispute resolution can only get harder. The
only feasible way forward is by getting the liability right. Banks which
destroy evidence should become automatically liable for the full sums in
dispute, including costs. Above all, the burden of proof must lie on the banks,
not the customer. The Payment Services Directive already requires this, yet
dispute resolution bodies like the UK Financial Ombudsman Service routinely
ignore the law and find for banks who destroy evidence.

\subsection{Industry Response}

We disclosed this vulnerability to the major card schemes and to selected banks
and payment switches in early 2012. All parties acknowledged receipt and
several contacted us to ask further questions. The card schemes chose initially
not to circulate the work, but after several weeks a different contact did
decide to circulate our report and our vulnerability disclosure report received
several thousand downloads. The vast majority of contacts refused to talk to us
on-the-record.

We received some informal responses: the extent and size of the problem was a
surprise to some, whereas others reported already being suspicious of the
strength of unpredictable numbers or even said others had been explicitly aware
of the problem for a number of years. If these assertions are true, it is
further evidence that banks systematically suppress information about known
vulnerabilities, with the result that fraud victims continue to be denied
refunds.

\section{Conclusions}
\label{sec:conclusions}

EMV is the main protocol used worldwide for card payments, being near universal
in Europe, in the process of adoption in Asia, and in its early stages in North
America. It has been deployed for ten years and over a billion cards are in
issue. Yet it is only now starting to come under proper scrutiny from
academics, media and industry alike. Again and again, customers have complained
of fraud and been told by the banks that as EMV is secure, they must be mistaken
or lying when they dispute card transactions. Again and again, the banks have 
turned out to be wrong. One vulnerability after another has been discovered and
exploited by criminals, and it has mostly been left to independent security
researchers to find out  what's happening and publicise it.

In this paper, we report the shocking fact that many ATMs and point-of-sale
terminals have seriously defective random number generators. These are often
just counters, and in fact the EMV specification encourages this by requiring
only that four successive values of a terminal's ``unpredictable number'' have
to be different for it to pass conformance testing. The result is that a crook
with transient access to a payment card (such as the programmer of a terminal
in a Mafia-owned shop) can harvest authentication codes which enable a
``clone'' of the card to be used later in ATMs and elsewhere.

The ``pre-play attack'' that we describe is not limited to terminals with
defective random number generators. Because of the lack of end-to-end
transaction authentication, it is possible to modify a transaction made with a
precomputed authentication code, en route from the terminal to the acquiring
bank, to edit the ``unpredictable number'' to the value that was used in the
pre-computation. This means that as well as inserting a man-in-the-middle
devices between the payment card and the terminal, an attacker could insert one
between the terminal and the acquirer. It also means that malware in the 
terminal can attack the EMV protocol even if the protocol itself is implemented
in a tamper-resistant module that the malware cannot penetrate. This may have
implications for terminals based on mobile phones that rely on cryptography in
the SIM card.

This flaw challenges current thinking about authentication. Existing models of
verification don't easily apply to a complex multi-stakeholder environment;
indeed, EMV has already been verified to be secure. We explained why such
verifications don't work and discussed the sort of analysis that is required
instead. Ultimately we feel that the tools needed to build robust systems for
millions of mutually mistrustful and occasionally hostile parties will involve
game-theoretic analysis as well as protocol-theoretic modelling. In addition,
mechanisms for rolling out fixes across networks with huge installed bases of
cards and terminals, and strong externalities, will have to be much better than
those we have at present, with incentives that put the pain where it's deserved
and technical mechanisms that offer the prospect of remedial action to the
sufferers.

In the meantime, there is a structural governance failure that gives rise to
systemic risk. Just as the world's bank regulators were gullible in the years
up to 2008 in accepting the banking industry's assurances about its credit risk
management, so also have regulators been credulous in accepting industry
assurances about operational risk management. In a multi-party world where not
even the largest card-issuing bank or acquirer or scheme operator has the power
to fix a problem unilaterally, we cannot continue to rely on a slow and complex
negotiation process between merchants, banks and vendors. It is time for bank
regulators to take an interest. It's welcome that the US Federal Reserve is now
paying attention, and time for European regulators to follow suit.

\subsection*{Acknowledgments}

The authors thank Dan Bernstein for the photo in \prettyref{fig:triton-desboard}. Steven J. Murdoch is funded by The Tor Project.

{\footnotesize \bibliographystyle{acm}
\bibliography{unattack}}

\end{document}